\documentclass[prb,showpacs,twocolumn]{revtex4}
\usepackage{graphicx}

\begin{document}


\title{Morphotropic  interfaces  in Pb(Mg$_{1/3}$Nb$_{2/3}$)O$_3$-PbTiO$_3$ single crystals}

\author{I. Rafalovskyi$^{\rm a}$}
\author{M. Guennou$^{\rm a}$}
\author{I. Gregora$^{\rm a}$}
\author{J. Hlinka$^{\rm a}$}
\email{hlinka@fzu.cz}
\affiliation{
$^{\rm a}$Institute of Physics, Academy of Sciences of the Czech Republic\\%
Na Slovance 2, 182 21 Prague 8, Czech Republic\\}

\date{\today}

\begin{abstract}
 The paper describes heterostructures spontaneously formed in (1-$x$)Pb(Mg$_{1/3}$Nb$_{2/3}$)O$_3$-$x$PbTiO$_3$ (PMN-xPT)
     single crystals cooled under bias electric field applied along [001]$_{\rm pc}$ and then zero-field-heated in the vicinity of the so-called
     depoling temperature  $T_{\rm RT}$. In particular,
      formation of lamellar structures composed of tetragonal-like and rhombohedral-like layers extending over
     macroscopic (mm) lengths
     is demonstrated
     by optical observations and polarized Raman investigations.
\end{abstract}

\pacs{77.80.-e,77.80.Dj,68.37.Ps}


\maketitle

      Large electromechanical coupling constant, piezoelectric coefficient and strain level
      of poled Pb(Mg$_{1/3}$Nb$_{2/3}$)$_{1-x}$Ti$_x$O$_3$
      (PMN-$x$PT) and
     similar single crystals have attracted a considerable interest because of their excellent performance
     in various solid state electromechanical sensors and
     actuators.\cite{Par97,Dam03,Zha12} It is well understood that the piezoelectricity
     results from the ferroelectric ordering and that the best piezoelectric figures of
     merit are found in materials with compositions at the so-called
     morphotropic phase boundary ($x\approx .33$) -  a  boundary
     separating stability domains of rhombohedral-like titanium-poor phase ($x \lesssim .33$)
     from the tetragonal-like titanium-rich ($x \gtrsim .33$) phase in the temperature-concentration phase diagram of the
     material.\cite{Zha12} This phase boundary is only very weakly temperature
     dependent, but in a narrow concentration region around $x \approx
     .33$ (which precisely comprises the materials of technological interest),  one may typically pass from the rhombohedral-like phase to
     the tetragonal-like phase at a certain\cite{Feng04} temperature  $T_{\rm RT}$, i.e. one may cross the MPB also upon
     heating.

     In these materials, the rhombohedral-like phase
     has in fact a very complicated micro and nano-scale domain texture.
     In fact,
     it is usually described  as a  lower symmetry phase, most often as monoclinic $Cm$ (also denoted as $M_{\rm A}$), or even as
     a fine mixture of several phases of different symmetry.\cite{Zha12,Jin03,Kam06,Chi06,Han03, comment1}

     Typically, to achieve high piezoelectric figures of
     merit, one uses rhombohedral-like material poled in a strong
     electric field applied along 100-direction (i.e., by a frustrative\cite{Ond09,Par97} poling which favors several domain states at a time).
      When such poled single crystals are heated above the $T_{\rm RT}$ temperature,   their ferroelecric domain structure
imposed by the frustrative poling is destroyed, and their
macroscopic piezoelectric properties are degraded. Therefore,
numerous efforts have been undertaken to increase of the $T_{\rm
RT}$ temperature (often denoted as {\it depoling}
temperature).\cite{Zha12}

Here we have investigated the passage across the $T_{\rm RT}$
temperature in a PMN-0.32PT single crystal, and, in the course of
these investigations, we have observed
  formation of {\it macroscopic} planar interfaces
  between the rhombohedral-like and tetragonal-like phase, or, in other words,
   interfaces that could be denoted as "real-space morphotropic phase boundaries".
    The optical observation has been complemented by local-probe polarized Raman spectroscopy investigations what
     allowed us to distinguish between the two phases.

The sample was a 0.5\,mm thick platelet of PMN-0.32PT single
crystal\cite{purchased} with main facets perpendicular to the
[100]$_{\rm pc}$ direction. These facets were covered by
evaporated gold electrodes. Optical observations and Raman
scattering experiments were performed in a reflection geometry
from an optically polished (001)$_{\rm pc}$ facet (perpendicular
to the principal (100)$_{\rm pc}$ facet, see Fig.\,1b).

\begin{figure}[ht]
\includegraphics[width=75mm]{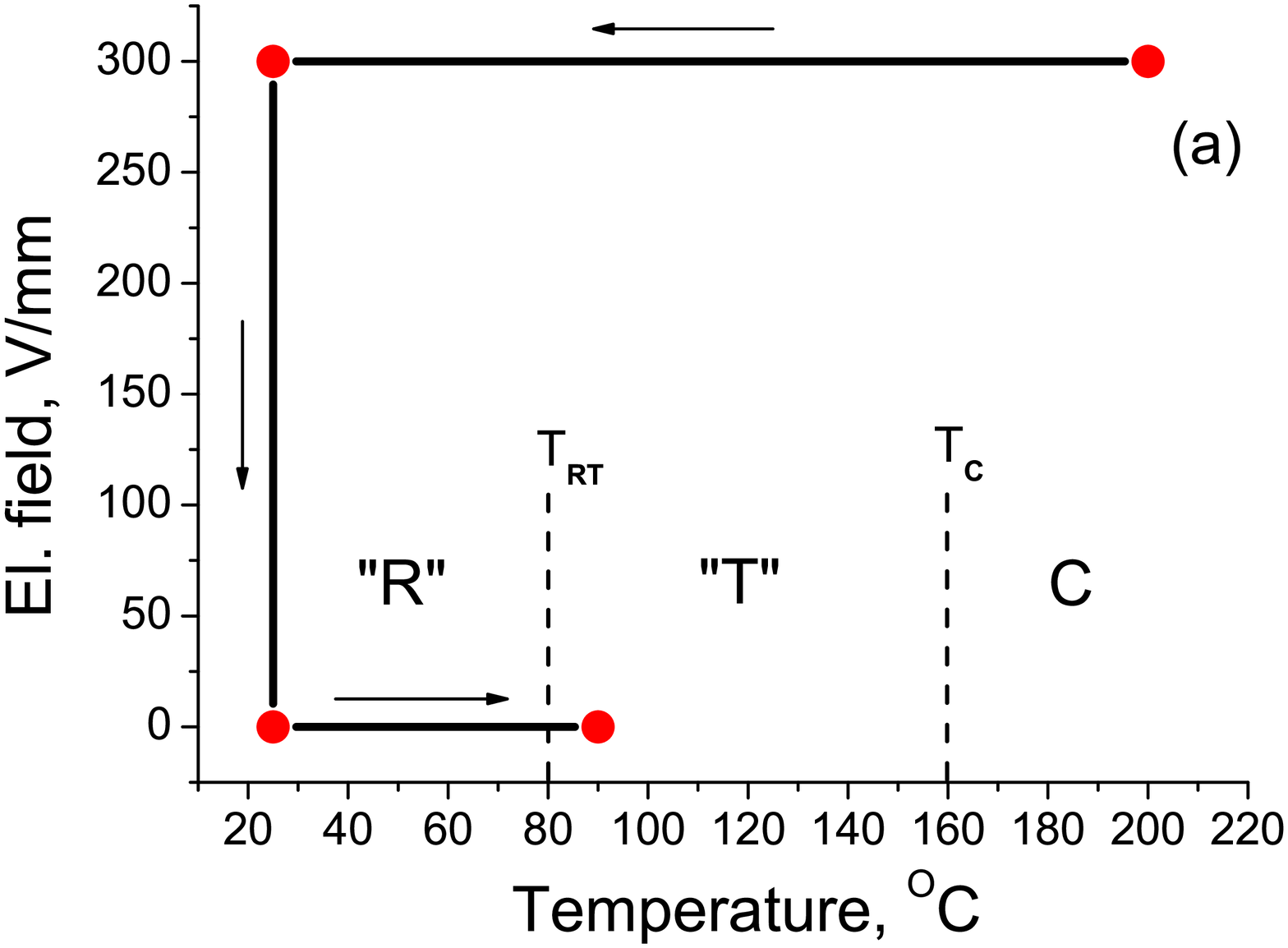}
\includegraphics[width=40mm]{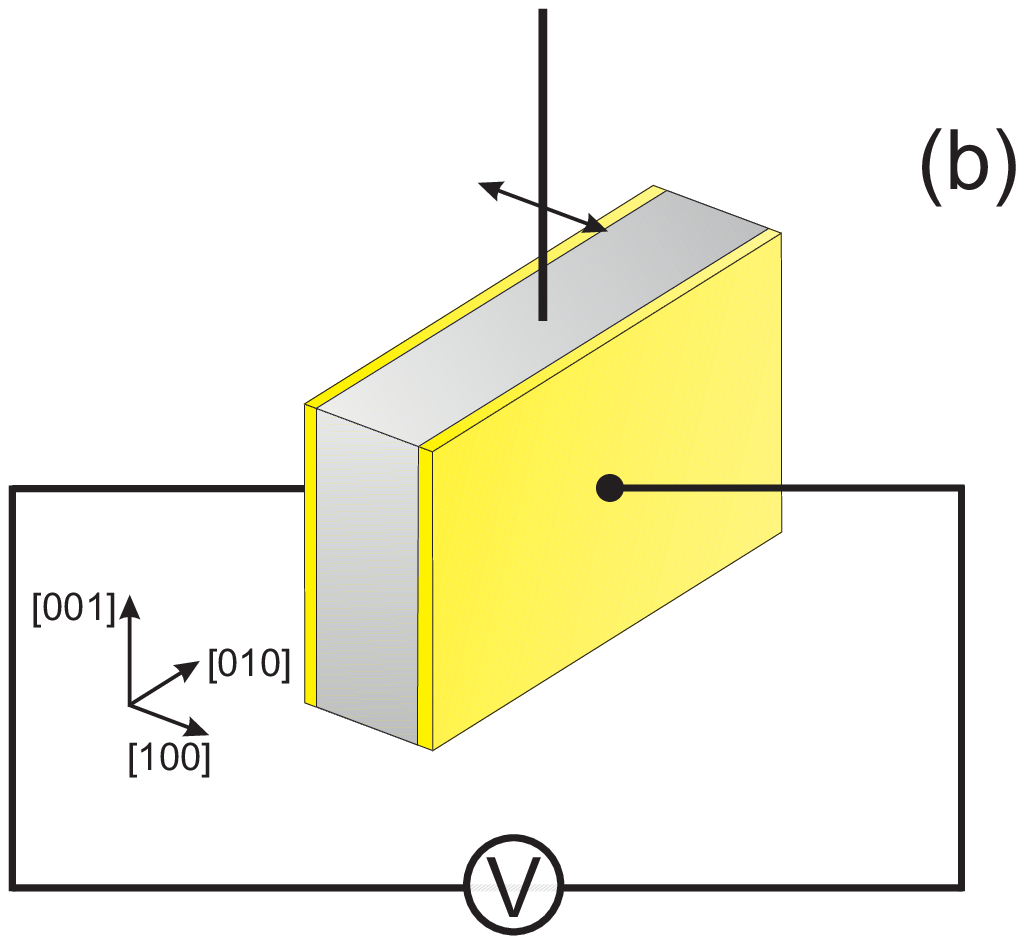}
 \caption{(Color online) Experimental arrangements used to prepare the laminate structure.
  (a) Schematic temperature vs [100] electric-field phase diagram with thermodynamic trajectory showing the adopted ZFHaFC experimental protocol,
  (b) sketch of the experimental geometry allowing in-situ Raman scattering or optical microscopy imaging of
  a facet perpendicular to the applied electric field.}
\label{figRafal1}
\end{figure}

For Raman scattering investigations, we have used a Renishaw RM
1000 micro-Raman spectrometer with 514.5 nm Ar laser excitation
line. Samples were placed into Linkam TS 1200 high temperature
cell where they were heated above the phase transition
temperature. The design of the cell allows applying electrical
field to the sample, so both electrical field and the temperature
were controlled during the experiment. The cell was mounted on the
rotation microscope table, and the sample position with respect to
polarization of incident light could be manually set to the
desired angle. With this experimental setup, we could use the same
microscope objective for Raman scattering as well as for optical
imaging {\it in situ}.

\begin{figure}
    \includegraphics[width=70mm]{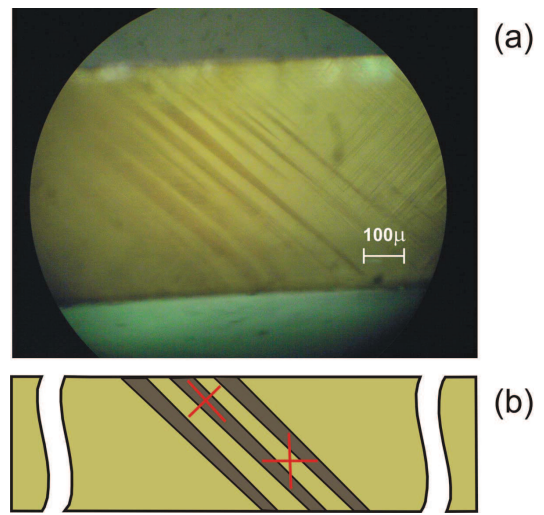}
 \caption{(Color online) Mesoscopic stripe pattern
 formed in a PMN-0.32PT crystal after the ZFHaFC process sketched in Fig.\,1.
  (a) optical micrograph in a reflection mode, (b) schematic illustration.
  Red crosses indicate orientation of polarizers on incident and scattered beam in configurations with a minimum Raman
  intensity at 570\,cm$^{-1}$.
  } \label{figRafal2}
\end{figure}

Prior to the Raman scattering measurement, the sample was field-cooled (5K\,/min)
from the annealing temperature of about 470\,K under a bias
electric field of about 300\,V/mm.
This thermal treatment was done with a sample already mounted in the optical cell,
 but up to this point we did not observed any obvious domain contrast in our experimental arrangement.
Then the electric field was
removed and the sample has been driven to about 360\,K (again at
about 5K\,/min). At this temperature, we could repeatedly observe
formation of a system of roughly parallel, 10-100 micron thick
stripes running across the sample at about 45 degrees to the edges
of the observed facet, as shown in Fig.\,\ref{figRafal2}.

At a first sight, these images are strongly reminiscent of the
observations of ferroelastic domains,\cite{Book2} but we shall
argue that here, in reality, the light and dark areas correspond
alternatively to lamellae of tetragonal-like and rhombohedral-like
domains, respectively. Obviously, the "morphotropic interfaces"
between such areas are similar to the usual ferroelastic domain
boundaries. In particular, the elongated wedge shape of the
observed interfaces indicates that the interfaces separate areas
with different spontaneous strain tensors, what may happen at the
twin boundary as well as at a two-phase interphase. However, we
have found that these stripes are formed only in the vicinity of
the anticipated $T_{\rm RT}$ temperature for this
composition\cite{Feng04,Zha12,Pera12,McLa05} (see Fig.\,1a).

In order to confirm this conjecture, we have investigated the
nature of these stripes by polarized Raman spectroscopy. In
general, it is known that Raman spectra of these materials change
only slightly at ferroelectric phase transitions. In fact, the
positions and shapes of the principal Raman bands in our
experiments were very similar to those reported in previous
investigations of PMN-$x$PT with a comparable
composition\cite{Slod10,Mar98,Kam03,Cha04,She05,Lim09,Slo08,Ohw01,Siny99}.
Fortunately, at least the relative intensities of Raman spectra
taken in dark and light stripes are different. In the following,
we will report Raman spectra in the cross-polarized (HV) geometry,
for which the differences were most apparent.

\begin{figure}[ht]
\includegraphics[width=57mm]{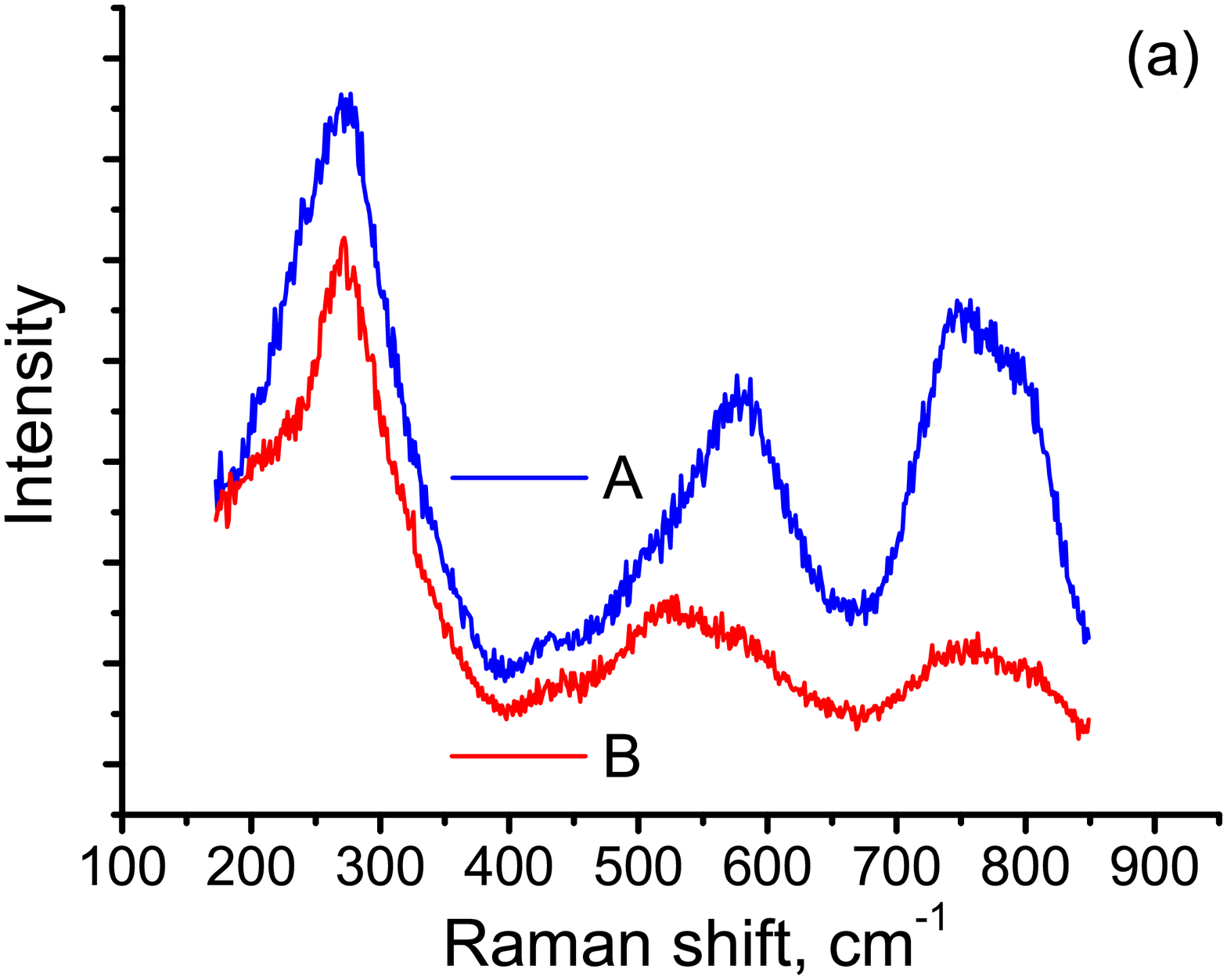}
\includegraphics[width=28mm]{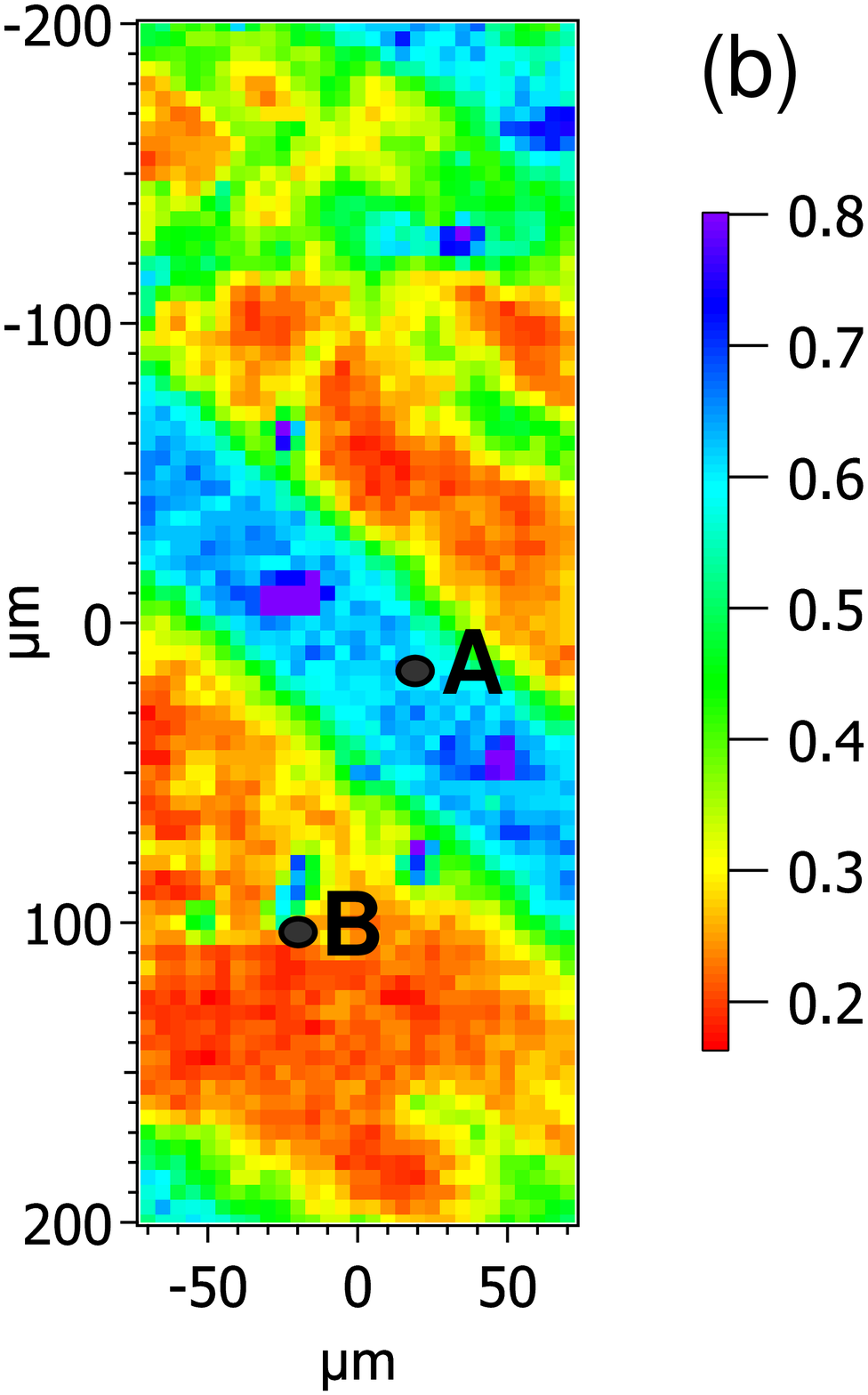}
 \caption{(Color online) Micro-Raman study  of the stripe pattern of PMN-0.32PT single crystal prepared by ZFHaFC process sketched in Fig.1a. (a)
Cross-polarized $[z(xy)z]$ Raman spectrum of PMN-0.32PT single
crystal taken from spots (A) and (B) within dark and light
stripes, respectively. (b) Map of the $I_{780}/I_{270}$ ratio of
the cross-polarized Raman scattering intensities recorded at
780\,cm$^{-1}$ and 270\,cm$^{-1}$ reveals the stripe pattern. (The
edges of the imaged area are parallel to the pseudocubic
 crystal axes and to the sample edges as well.) } \label{figRafal3}
\end{figure}

 Spectra from selected dark and light stripes are
shown in Fig.\,\ref{figRafal3}a. The difference in intensity is
most obvious when comparing the shape and overall
  intensity of the phonon bands near 600\,cm$^{-1}$ or
  800\,cm$^{-1}$. The difference can be quantified using ratios of intensities between different bands. For example, the ratio $I_{570}/I_{510}$ of the cross-polarized Raman
  intensity
  detected at 570\,cm$^{-1}$ with respect to that detected at 510\,cm$^{-1}$ varies by more than 50 percent when comparing dark and light
  stripes, and similar contrast is obtained when considering the ratio $I_{780}/I_{270}$ of intensities
  at 780\,cm$^{-1}$ and 270\,cm$^{-1}$. Note that both these ratios were previously used in
  Ref.\,\onlinecite{Yang10} for distinguish two types of coexisting structural "microregions" of irregular shape in PMN-0.33PT, assigned to the $M_A$ ({\it i.e.} rhombohedral-like phase) and $M_C$ ({\it i.e.} tetragonal-like phase).
  We used the latter ratio to map the stripes and observed that in contrast to results
  of Ref.\,\onlinecite{Yang10}, in our sample the so-obtained relative intensity map
  reveals clear, about $100$-micron sized stripes, which clearly
match the stripes observed optically (Fig.\,\ref{figRafal3}b).

\begin{figure}[ht]
  \includegraphics[width=65mm]{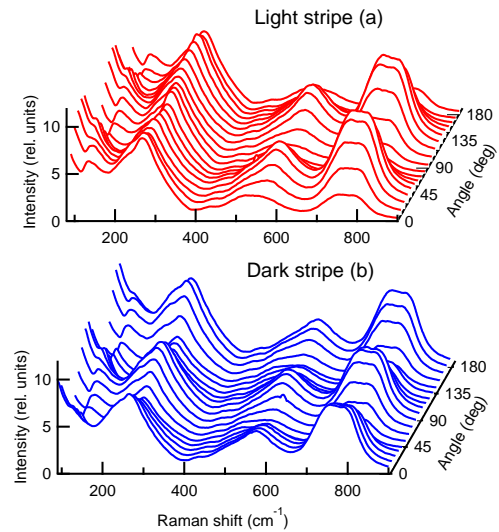}
 \caption{(Color online) Cross-polarized backscattering Raman spectra of PMN-0.32PT single crystal
 prepared by ZFHaFC process sketched in Fig.1\,a, taken
 from a light (a) and dark (b) stripe, respectively, at varying positions of the
 polarizer axis (the indicated angle $\phi$ defines the direction of polarizer axis with respect to pseudocubic axes of the crystal - for $\phi=0$,
  the polarizers are parallel to the pseudocubic
 crystal axes and sample edges). The angular dependence suggests that the light and dark
 stripes correspond to a tetragonal-like and rhombohedral-like ferroelectric structural variants, respectively.}
 \label{figRafal4}
\end{figure}

To clarify the origin of the Raman contrast associated with the
stripes observed, we have recorded the variation of the Raman
spectra in a given stripe as a function of the angle $\phi$
between the polarizer and pseudocubic crystal axes, parallel to
the edges of the sample. Two sets of spectra collected from laser
spot focused within dark and light stripe, respectively, are shown
in Fig.~\ref{figRafal4}. The angular dependence of Raman intensity
recorded in the dark area is typical for the rhombohedral-like
(relaxor) phase of PMN-$x$PT crystals (with maxima at $\phi=
n\,\pi/2, n=0,1,2,...$), and it also agrees with the measurements
of pure PMN, documented for example in Ref.\,\onlinecite{Tani11}.
In contrast, the angular dependence of Raman intensity recorded in
the brighter areas shows the opposite behavior - positions of the
intensity maxima are shifted by about 45 degrees with respect to
that of the dark areas.

Such angular shift cannot be simply explained by measurement in a
different domain of the same rhombohedral phase: all possible
Raman tensors of inequivalent rhombohedral ferroelastic domain
states can be obtained by rotation around the fourfold axis
parallel to the [001] direction. Also, the angular dependence of
Raman intensity in elastically compatible tetragonal domain pair
is not compatible with our observation. Similar situation can be
expected for monoclinic phases that are close to either tetragonal
or rhombohedral phase only. On the contrary, projection of the
optical axes of one rhombohedral and one tetragonal domain are
mutually at 45 degrees. Therefore, the observed angular shift
corroborates well the anticipated two-phase picture.

The available experimental setup did not allow a precise in-situ
determination of the observed interphases, but the optical
observations hint that the interfaces are close to the $\{110\}$
crystallographic planes. The repeating sequence of nearly parallel
interfaces suggests that these are coherent interfaces similar to
twin boundaries and that their orientation allows to minimize the
interface stresses. Coexistence of different phases is rather
frequent phenomenon in case of first-order phase transformations
and a planar macroscopic phase front can be stabilized for example
with the aid of temperature or concentration
gradient.\cite{Born99} Macroscopic phase boundary between
different ferroelectric phases has been previously observed in
PMN-$x$PT crystal with a linear gradient of Ti
concentration\cite{Shu05}. However, in that work, the phase
boundary seems to simply delimit Ti-poorer regions from Ti-richer
regions, without any obvious relation to the crystallographic
orientation of the sample.
 Much closer to
our observation is the figure 7 of Ref.\,\onlinecite{Fuj98},
showing coexistence of rhombohedral and tetragonal phase in a
single crystal of Pb(Zn$_{1/3}$Nb$_{2/3}$)$_{1-x}$Ti$_x$O$_3$
(PZN-$x$PT). There also the stripes of the tetragonal-like phase
are intercalated within the rhombohedral phase and oriented at
about 45 degrees with respect to [001]$_{\rm pc}$
direction.\cite{comment2} Thus, the phenomenon observed here seems
to be common to both PMN-$x$PT and PZN-$x$PT systems.

In fact, the angular dependence recorded in the dark and light
areas here are also quite analogous to that of "G microregions"
and "R microregions", respectively, discovered in PMN-0.33PT
single crystals by authors of the Ref.\,\onlinecite{Yang10} and
then also reinvestigated in the temperature study of
Ref.\,\onlinecite{Yan11}. In contrast with our observations, the
borders of the previously observed "G microregions" and "R
microregions" have an irregular, spongy character. In the light of
the present observations it seems quite likely that these
microregions do correspond to the rhombohedral and tetragonal-like
areas, even though the geometry of the interfaces is somewhat
unexpected. Different morphology is perhaps related to some frozen
defects or concentration fluctuations. We also note that their
PMN-0.33PT sample was studied "as-grown", whereas our observations
were made with PMN-0.32PT sample and the macroscopic interfaces
were only observed after poling of the sample.

At the microscopic level, coexistence of the rhombohedral-like and
tetragonal like phases in relaxor materials with composition close
to the phase boundary has been reported in a number of recent
works, and it is possible that such interfaces could play some key
role in their high piezoelectric
properties\cite{Wu12,Sato11,Schm10,Ma12,Iwa12}, even though for
example the microscopic mechanisms responsible for the
piezoelectricity in Na$_{1/2}$Bi$_{1/2}$TiO$_{3}$ based materials
are very different from that of lead-based materials.\cite{GuenUP}
 For example, it is possible that the
phenomenon of overpoling of lead-based relaxor ferroelectric materials
 is related to the presence of
{\em macroscopic} (percolated) tetragonal-like areas, while {\em microscopic}
phase coexistence could still favor the piezoelectric
response.\cite{Raja07}
 In either case, we believe that further studies of the
macroscopic phase interfaces could help to understand the
functional properties of this family of materials.

Our experimental observations clearly indicate that a slow
zero-field heating of previously field-cooled PMN-0.32PT specimens
is a protocol suitable for the stabilization of macroscopic
interfaces between tetragonal and rhombohedral-like (possibly
monoclinic) variants of PMN-$x$PT single crystals.
 Formation of tetragonal and rhombohedral-like stripes occurs in near the $T_{\rm RT}$ transition.

Because of their macroscopic (millimetric) size, these
"morphotropic interfaces" separating adjacent stripes can be
easily observed in an optical microscope. In fact, the existence
of macroscopic interfaces allowed us also to employ spatially
resolved polarized Raman scattering techniques to confirm that the
stripe pattern is due to the coexistence of the phases attached to
the opposite sides of the morphotropic phase boundary in the
temperature-composition phase diagram. However, we believe that
the existence of such macroscopic interfaces can be advantageous
for other investigations of the coexistence and compatibility of
tetragonal and rhombohedral-like (possibly monoclinic) variants
MPB systems. Our results also suggest that MPB materials have a
clear potential to sustain different interfaces than just
"ordinary" ferroelectric domain walls separating two domains with
the same ferroelectric phase. This might considerably enrich the
domain-wall engineering opportunities available for this family of
MPB crystals.

{\em Acknowledgment.} This work was supported by the Czech Science
Foundation (Projects Project GACR P204/10/0616
 and 202/09/H0041). In
addition, the contribution of Ph.D. student I. Rafalovskyi has
been supported by Czech Ministry of Education (project
SVV-2013-267303).


\begin{thebibliography}{99}
\bibitem{comment2} {According to the indicated extinction condition, it seems that
the investigated PZN-$x$PT sample was viewed along (001)$_{\rm
pc}$ direction.}
\end{thebibliography}
\end{document}